\documentclass[twocolumn, pra, showpacs,superscriptaddress]{revtex4}
\usepackage{graphicx}
\usepackage{dcolumn}
\usepackage{bm}
\usepackage{amsmath}

\begin{document}

\title{Density distributions for trapped one-dimensional spinor gases}
\author{Yajiang Hao}
\affiliation{Department of Physics and Institute of Theoretical
Physics, Shanxi University, Taiyuan 030006, P. R. China}
\affiliation {Institute of Physics, Chinese Academy of Sciences,
Beijing 100080, P. R. China}
\author{Yunbo Zhang}
\affiliation{Department of Physics and Institute of Theoretical
Physics, Shanxi University, Taiyuan 030006, P. R. China}
\author{J. Q. Liang}
\affiliation{Department of Physics and Institute of Theoretical
Physics, Shanxi University, Taiyuan 030006, P. R. China}
\author{Shu Chen}
\email{schen@aphy.iphy.ac.cn}
\affiliation{Institute of Physics,
Chinese Academy of Sciences, Beijing 100080, P. R. China}

\begin{abstract}
We numerically evaluate the density distribution of a spin-1
bosonic condensate in its ground state within a modified
Gross-Pitaevskii theory, which is obtained by the combination of
the exact solution of the corresponding integrable model with the
local density approximation. Our study reveals that atoms in the
$m_F=0$ state are almost completely suppressed for the
anti-ferromagnetic interactions in both weakly and strongly
interacting regimes, whereas all three components remain
non-vanishing for ferromagnetic interactions. Specially, when the
system is in the Tonks-Girardeau (TG) regime, obvious Fermi-like
distribution emerges for each component. We also discuss the
possible deviation of the spatial distribution from the Fermi-like
distribution when the spin-spin interaction is strong enough.
\end{abstract}
\maketitle



\narrowtext

\section{Introduction}

The experimental realization of Bose-Einstein condensates (BECs)
of trapped alkali atomic clouds \cite{Anderson} has stimulated
active studies in many new regimes. From then on, BECs have become
a popularly investigated platform for various effects of quantum
many-body interaction in strongly correlated systems. When BECs
are realized in optical traps rather than in magnetic ones, the
spin degrees of freedom in alkali atoms are liberated. As a
consequence, such a system with internal degrees of freedom
manifests very rich physics and many fascinating phenomena have
been observed in spinor condensates, e.g., quantum entanglement of
spins, spinor four-wave mixing, and spin domains, etc.
\cite{Andersen,T.L.Ho,T.L.Ho2,Law,Jap,WP Zhang}.

On the other hand, in the research area of low-dimensional physics,
the effect of dimensionality reduction in bosonic system has been
investigated extensively \cite
{Olshanii,Dunjko,Luxat,Pedri,Chen,Kunal,ohberg,Girardeau1} and is
being paid more and more attention \cite{Recati,Astrakharchik}.
There has been tremendous experimental progress towards the
realization of trapped one-dimensional (1D) cold atom systems \cite
{gorlitz,esslinger,richard,YJWang}. Very recently, several groups
have reported the observation of a 1D Tonks-Girardeau (TG) gas \cite
{Paredes,Kinoshita}. An array of 1D quantum gas is obtained by
tightly confining the particle motion in two directions to zero
point oscillations \cite{Ketterler} by means of two-dimensional
optical lattice potentials. By loading the condensate in optical
lattice and changing the trap intensities, and hence the atomic
interaction strength, the atoms can be made to act either like a
condensate or like a TG gas. The TG gas provides a textbook example
where atom-atom interaction plays a critical role and mean-field
theory fails to obtain reasonable results \cite{Girardeau2}. The
dimensionless interaction parameter $\gamma =mg/\hbar ^{2}\rho $
governs the crossover from the weakly interacting condensate to the
strongly interacting TG gas, where $g$ is an effective 1D
interaction constant, $m$ is the mass of the atom, and $\rho $ is
the 1D density of the quantum gas.

Within the mean field theory, Zhang and You recently studied the
spin-1 atomic condensate in a cigar-shaped trap by solving an
effective quasi-1D nonpolynomial Schr\"{o}dinger equation \cite{WX
Zhang}. As is well known, the mean field theory fails to work if
the system enters the strongly interacting TG regime
\cite{Kolomeisky}. It is therefore desirable to investigate the 1D
spinor gases within a theoretical approach valid to the strongly
interacting TG regime as well as the weakly interacting
Thomas-Fermi regime. In this paper, we investigate the ground
state properties of the 1D spinor Bose gases under the local
density approximation \cite{Dunjko,Kolomeisky,ohberg} combining
with the exact result of the solvable Bose gas model
\cite{Lieb,Sutherland,Li}. The dependence of density profile on
the magnetization and the effects of many-body interaction are
studied in both the weakly interacting regime ($\gamma \ll 1$) and
the TG regime ($\gamma \gg 1$). The density profile for the 1D
spinor Bose gases in the ground states is then obtained by
numerically solving the modified Gross-Pitaevskii Equations
(GPEs). Particularly, when the system is in the TG regime, we
found that the effect of spin-spin interaction could be dramatic.

The paper is organized as follows. In Sec. II, we give a brief
review of the 1D spinor model trapped in an external potential and
derive the modified GPEs. In Sec. III, we introduce our numerical
procedure for solving the coupled GPEs. Aiming at the realistic
systems which may be accessible in experiments, in Sec IV, we
present results of the ground state density distributions for the
trapped spinor gases in different interacting regimes. A brief
summary is given in Sec. V.

\section{Formulation of the model and method}

The $s$-wave scattering between two identical spin-1 bosons is
characterized by the total spin of two colliding bosons, $0$ or
$2$, and we denote the corresponding scattering lengths by $a_0$
and $a_2$. When the cold atoms were trapped intensively in
transverse direction with the transverse trapping frequency $\hbar
\omega _{\bot }$ greatly exceeding the chemical potential, the
radial motion of atoms is essentially `frozen out'. The dynamics
is thus governed by a 1D Hamiltonian with the effective 1D
interaction strength given by \cite{Olshanii,Olshanii2,Petrov}
\begin{eqnarray}
U_{0,2} &=&-\frac{2\hbar ^2}{ma_{0,2}^{1D}}, \\
a_{0,2}^{1D} &=&-\frac{d_{\bot }^2}{2a_{0,2}}\left(
1-\mathcal{C}\left( a_{0,2}/d_{\bot }\right) \right),   \nonumber
\end{eqnarray}
where $d_{\bot }=\sqrt{\hbar /m\omega _{\bot }}$ and $\mathcal{C}\approx
1.4603$. For arbitrary hyperfine states in the hyperfine manifold in which
there are two spin-$1$ atoms, the effective interaction may therefore be
written as
\begin{eqnarray}
U_{ij} &=&\delta \left( x_i-x_j\right) \left(
U_0\mathcal{P}_0+U_2\mathcal{P}
_2\right)  \\
&=&\delta \left( x_i-x_j\right) \left( c_0+c_2\mathbf{F}_i\cdot
\mathbf{F} _j\right) ,  \nonumber
\end{eqnarray}
where the operators $\mathcal{P}_0$ and $\mathcal{P}_2$ project the
wave function of a pair of atoms into a state of total spin $0$ and
$2$. The coefficients $c_0$ and $c_2$ are related to the effective
1D interaction constants $U_0$ and $U_2$ through
$c_0=\frac{U_0+2U_2}3$ and $c_2=\frac{U_2-U_0}3$. The ground state
of the system is ferromagnetic if $c_2<0$, and anti-ferromagnetic if
$c_2>0$. $F_x$ , $F_y$ and $F_z$ are three spin-1 matrices with the
quantization axis taken along the $z$-axis direction.

For a spin-1 Bose condensate trapped in an external potential
$V_{ext}\left( x\right) $ which is independent of the internal
states, the Hamiltonian can be expressed in the second quantized
form as
\begin{eqnarray}
\mathcal{H} &=&\int dx\hat{\Psi}_i^{+}\left( -\frac{\hbar ^2}{2m}\frac{d^2}{%
dx^2}+V_{ext}\left( x\right) \right) \hat{\Psi}_i  \nonumber \\
&&+\frac{c_0}2\int dx:\left( \hat{\Psi}_i^{+}\hat{\Psi}_i\right) ^2:
\nonumber \\
&&+\frac{c_2}2\int dx\hat{\Psi}_k^{+}\hat{\Psi}_i^{+}\left( F_\eta \right)
_{ij}\left( F_\eta \right) _{kl}\hat{\Psi}_j\hat{\Psi}_l,  \label{Hspinor}
\end{eqnarray}
where $\hat{\Psi}_i\left( x\right) $ ($\hat{\Psi}_i^{\dag }\left(
x\right) $ ) is the field operator that annihilates (creates) an
atom in the $i$-th internal state at location $x$, and $i=+,0,-$
denotes the atomic hyperfine state $\left|
F=1,m_F=+1,0,-1\right\rangle $ respectively. Summation is assumed
for repeated indices in the above Hamiltonian and the pair of
colons denote the normal-order product. Within the mean field
approach the properties of a spinor gas are determined by the
following spin-dependent energy functional
\begin{eqnarray}
\mathcal{E} &=&\int dx\left[ \Phi _i^{*}\left( -\frac{\hbar
^2}{2m}\frac{d^2 }{dx^2}+V_{ext}\left( x\right) \right) \Phi
_i+\rho \epsilon \left( \rho
\right) \right]  \nonumber \\
&&+\int dx\left[ \frac{c_2}2\Phi _k^{*}\Phi _i^{*}\left( F_\eta \right)
_{ij}\left( F_\eta \right) _{kl}\Phi _j\Phi _l\right] ,  \label{EF}
\end{eqnarray}
where $\rho =\sum_i\rho _i=\sum_i\left| \Phi _i\right| ^2$ and
$\epsilon \left( \rho \right) =c_0\rho /2$.

For a one-component Bose gas, it turns out that such a mean field
approach works well in the weakly interacting regime but is not
able to describe the density distribution correctly in the TG
regime. However, a modified Gross-Pitaevskii theory, within which
the interaction effect is properly taken into account by using the
Lieb-Liniger solution, proves to be able to yield accurate
ground-state density distribution even in the Tonks limit
\cite{Dunjko,Chen,Kolomeisky,ohberg,Kim}. The modified
Gross-Pitaevskii theory essentially is based on the local density
approximation together with the analytical solution of the
homogenous system. Within the local density approximation, the
chemical potential of the inhomogeneous gas is determined by the
local equilibrium condition,
\[
\mu = \mu_{hom}[\rho(x)]+V_{ext}(x),
\]
with $\mu_{hom}[\rho]$ being the chemical potential of the
corresponding homogeneous system. For the one-component Bose gas,
$\mu_{hom}[\rho]$ can be obtained from the solution of the
Lieb-Liniger model. In our present problem, the spin exchange
strength $c_2$ is much smaller than the density-density
interaction $c_0$ and the model (\ref{Hspinor}) for $V_{ext}(x)=0$
and $c_2=0$ is the integrable three-component Bose gas
model\cite{Sutherland,Li}. To make progress, we work in the scheme
of modified Gross-Pitaevskii theory
\cite{Dunjko,Minguzzia,Kolomeisky} and take $\mu_{hom}[\rho]$ from
the corresponding integrable model of three-component Bose gas.
Effectively, this is equivalent to replacing the energy density
$\epsilon \left( \rho \right) $ in Eq. (\ref{EF}) with the energy
density of the exactly solvable three-component Bose model, which
takes the values in the two limiting cases as following
\begin{equation}
\epsilon \left( \rho \right) =\frac{\hbar ^2}{2m}\rho e\left( \gamma \right)
=\{
\begin{array}{ll}
\text{ \ \qquad }c_0\rho /2, & \gamma \ll 1, \\
\pi ^2\hbar ^2\rho ^2/6m, & \gamma \gg 1.
\end{array}
\end{equation}
The ground state energy density of the three-component Bose model
has a similar form as its one-component correspondence, i.e., the
Lieb-Liniger model, however, here the density $\rho=\rho_{+}
+\rho_{0}+\rho_{-}$ is the total density of three components. We
note that in Eq. (\ref{EF}) the kinetic energy term is associated
with the inhomogeneity of the gas due to external confinement
$V_{ext}\left( x\right)$\cite{ohberg,Lieb2003} and $\epsilon
\left( \rho \right) $ represents essentially the energy density in
the homogeneous system, while the last term the spin-spin
interaction energy in 1D.

In the weakly interacting regime ($\gamma \ll 1)$ the interaction
energy will not change the wave function greatly because it is
negligibly small compared with the characteristic kinetic energy of
an individual atom. In the TG regime ($\gamma \gg 1$), however, the
interaction between atoms is so strong that the bosonic atoms behave
much like spinless fermions. In both cases, the spin dependent term
can be expressed in the explicit form
\begin{eqnarray}
&&\Phi _k^{*}\Phi _i^{*}\left( F_\eta \right) _{ij}\left( F_\eta \right)
_{kl}\Phi _j\Phi _l  \nonumber \\
&=&\rho _{+}^2+\rho _{-}^2+2\rho _0\rho _{-}+2\rho _{+}\rho _0-2\rho
_{+}\rho _{-}  \nonumber \\
&&+2\Phi _0^{*2}\Phi _{+}\Phi _{-}+2\Phi _0^2\Phi _{-}^{*}\Phi _{+}^{*}.
\end{eqnarray}
We note that the last two terms in the above equation correspond
to the processes that would change the spin states. An atom in the
$ m_F=1$ state scatters with another atom in the $m_F=-1$ state,
and consequently it produces two atoms in the $m_F=0$ state or
vice versa. Nevertheless, these processes conserve the
magnetization of the system $\mathcal{M}=\int dx\left\langle
F\right\rangle =\int dx\left[ \Phi _{+}^{*}\Phi _{+}-\Phi
_{-}^{*}\Phi _{-}\right] $ \cite{S. Yi,WX Zhang}. In order to
obtain the ground state from a global minimization of
$\mathcal{E}$ with the constraints on both $N$ and $\mathcal{M}$,
we introduce separately Lagrange multiplier $B$ to conserve $M$
and the chemical potential $\mu $ to conserve $N$. The ground
state is then determined by a minimization of the free-energy
functional $\mathcal{F}= \mathcal{E}-\mu N-B\mathcal{M}$. The
dynamics of $\Phi _i$ is governed by the coupled GPEs
\begin{eqnarray}
i\hbar \partial \Phi _{+}/\partial t &=&\left[ H-B+c_2\left( \rho _{+}+\rho
_0-\rho _{-}\right) \right] \Phi _{+}+c_2\Phi _0^2\Phi _{-}^{*},  \nonumber
\\
i\hbar \partial \Phi _0/\partial t &=&\left[ H+c_2\left( \rho _{+}+\rho
_{-}\right) \right] \Phi _0+2c_2\Phi _{+}\Phi _{-}\Phi _0^{*}, \\
i\hbar \partial \Phi _{-}/\partial t &=&\left[ H+B+c_2\left( \rho _{-}+\rho
_0-\rho _{+}\right) \right] \Phi _{-}+c_2\Phi _0^2\Phi _{+}^{*},  \nonumber
\end{eqnarray}
with
\begin{equation}
H=-\frac{\hbar ^2}{2m}\frac{d^2}{dx^2}+V_{ext}\left( x\right)
+\tilde{F} \left( \rho \right)
\end{equation}
and
\begin{equation}
\tilde{F}\left( \rho \right) =\frac \partial {\partial \rho }\left[ \rho
\epsilon \left( \rho \right) \right] =\{
\begin{array}{ll}
\text{ \ \qquad }c_0\rho, & \gamma \ll 1, \\
\pi ^2\hbar ^2\rho ^2/2m, & \gamma \gg 1.
\end{array}
\end{equation}
By numerically solving the above equations, we will determine the
ground state density distributions for the 1D spinor Bose gases
trapped in an harmonic trap $V_{ext}\left( x\right) =\frac
12m\omega ^2x^2$ both in the weakly interacting regime and in the
TG regime.

\section{Numerical method}

To simplify the formalism we choose a slightly different notation.
Taking advantage of the fact that all distances and energies in
the calculation can be scaled in units of typical length and
energy of the external harmonic potential, we introduce the
standard length unit
\begin{equation}
a=\sqrt{\frac{\hbar }{m\omega }},
\end{equation}
and rescale the spatial coordinate, the wave function, and the time
variable as
\begin{eqnarray}
x &=&\tilde{x}a,  \nonumber \\
\Phi _i &=&\sqrt{\frac Na}\phi _i,  \nonumber \\
\tilde{\rho}_i &=&\left| \phi _i\right| ^2,  \nonumber \\
\tilde{\rho} &=&\sum_{i=1}^3\tilde{\rho}_i,  \nonumber \\
t &=&\frac \tau \omega .  \nonumber
\end{eqnarray}
The wave function is thus normalized to $\int d\tilde{x}\sum_{i=1}^3\left|
\phi _i\right| ^2=1$. With the above changes, our coupled nonlinear
Schr\"{o}dinger equations become
\begin{eqnarray}
i\partial \phi _{+}/\partial \tau &=&\left[ \tilde{H}-\tilde{B}+g_2\left(
\tilde{\rho}_{+}+\tilde{\rho}_0-\tilde{\rho}_{-}\right) \right] \phi
_{+}+g_2\phi _0^2\phi _{-}^{*},  \nonumber \\
i\partial \phi _0/\partial \tau &=&\left[ \tilde{H}+g_2\left( \tilde{\rho}%
_{+}+\tilde{\rho}_{-}\right) \right] \phi _0+2g_2\phi _{+}\phi _{-}\phi
_0^{*},  \label{GPEs2} \\
i\partial \phi _{-}/\partial \tau &=&\left[ \tilde{H}+\tilde{B}+g_2\left(
\tilde{\rho}_{-}+\tilde{\rho}_0-\tilde{\rho}_{+}\right) \right] \phi
_{-}+g_2\phi _0^2\phi _{+}^{*},  \nonumber
\end{eqnarray}
where $\tilde{B}=NB/a\hbar \omega $ and $\tilde{H}=-\left( 1/2\right) d^2/d%
\tilde{x}^2+\tilde{x}^2/2+\tilde{F}\left( \tilde{\rho}\right) $ with
\[
\tilde{F}\left( \rho \right) =\{
\begin{array}{ll}
g_0\tilde{\rho} & \gamma \ll 1, \\
\tilde{g}_0\tilde{\rho}^2 & \gamma \gg 1.
\end{array}
\]
The pair interaction constants are also rescaled as
\begin{eqnarray*}
g_0 &=&Nc_0/a\hbar \omega , \\
\tilde{g}_0 &=&N^2\pi ^2/2, \\
g_2 &=&Nc_2/a\hbar \omega .
\end{eqnarray*}

By propagating the coupled GPEs Eq. (\ref{GPEs2}) in imaginary
time, we obtain the ground state of spin-$1$ BECs in one
dimension. In each propagating step, the wave function $\phi _i$
is normalized to conserve the atomic number. We ensure the
conservation of magnetization $\mathcal{M}$ by adjusting the
Lagrange multiplier $B$. In our procedure the Crank-Nicholson
scheme is used to discretize Eq. (\ref{GPEs2}) \cite{Recipes}. We
take the initial wave function to be a complex Gaussian with a
constant velocity: $exp\left[ -x^2/2q-ikx\right] $. Here $q$ and
$k$ are adjustable parameters that shall not affect the final
converged ground state \cite{S. Yi}.
\begin{figure}[tbp]
\includegraphics[width=3.5in]{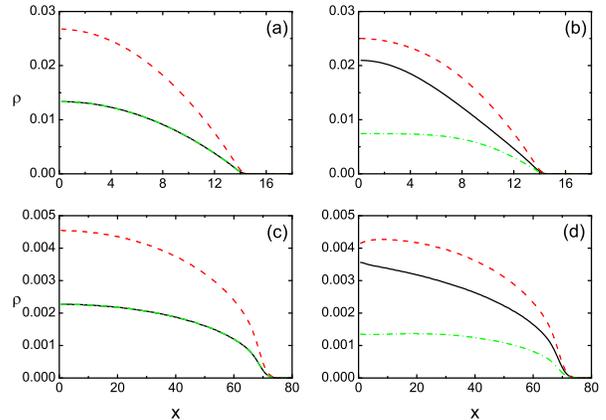}
\caption{(Color online) The density profiles of the spin-1
$^{87}$Rb condensates in the ground state for the + (solid line),
0 (dashed line) and - (dash-dot lines) components respectively.
(a) m=0, Thomas-Fermi regime; (b) m=0.2, Thomas-Fermi regime; (c)
m=0, Tonks regime; (d) m=0.2, Tonks regime. In this figure the
lengthes are in units of 1.2$\mu $m.} \label{fig1}
\end{figure}

\section{Density profiles in the ground state}

To give a concrete example, we firstly evaluate the density
profiles of 1D spinor gases in the ground state for $^{87}$Rb
(ferromagnetic) with $ a_0=102a_B$ and $a_2=100a_B$ ($a_B$ is the
Bohr radius) \cite{Rb}. By properly choosing the parameters, the
system may be prepared either in the weakly interacting regime or
in the TG regime. Let us first consider the specific system with
typical trap parameters $\omega _x=0.5$kHz and $ \omega _{\bot
}=50$kHz for $N=2000$ atoms, in which case the effective
interaction strength $\gamma \sim 0.008$ indicating that the
system is in the weakly interacting regime. Fig. 1(a) and Fig.
1(b) show the density profiles in units of $N/a$ for
m$=\mathcal{M}/N=0$ and m$=0.2$, respectively. The density
profiles of $+$ component (solid line) and $-$ component (dash-dot
lines) superpose on each other exactly as m$=0$. When the
parameters are tuned to $ \omega _x=10$ Hz, $\omega _{\bot
}=500$kHz and the atomic number to $N=50$, the effective
interaction strength $\gamma \sim 15$ indicates that the system is
in the TG regime. The corresponding density profiles are plotted
in Fig. 1(c) and Fig. 1(d) for m$=0$ and m$=0.2$, respectively. In
the TG regime, the atoms distribute uniformly in more extensive
area because of the strong interaction in the system. At the
boundary the density profiles decrease to zero rapidly, which
reminds us the Fermi-Dirac statistical distribution. This imply
that the density distribution of the bosonic atoms in the TG
regime behave like that of the Fermions.
\begin{figure}[tbp]
\includegraphics[width=3.5in]{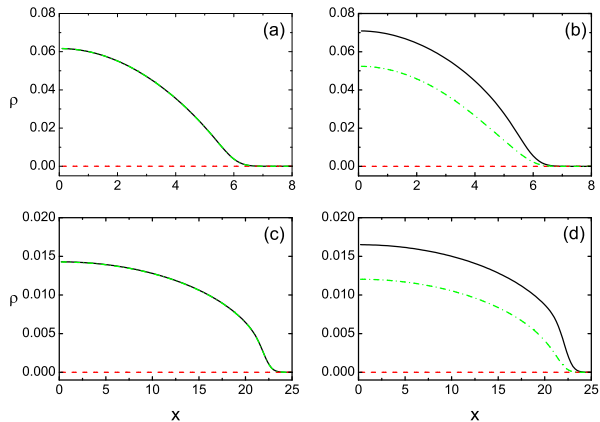}
\caption{(Color online) The density profiles of the spin-1
$^{23}$Na condensates in the ground state for the + (solid line),
0 (dashed line) and - (dash-dot lines) components respectively.
(a) m=0, Thomas-Fermi regime; (b) m=0.2, Thomas-Fermi regime; (c)
m=0, Tonks regime; (d) m=0.2, Tonks regime. In this figure the
lengthes are in units of 7.4$\mu $m.} \label{fig2}
\end{figure}

For the anti-ferromagnetic system we consider a condensate of
$^{23}$Na with$ \ a_0=50a_B$ and $a_2=55.1a_B$ \cite{Na}. In the
weakly interacting regime, the trap frequencies are chosen as
$\omega _x=50$Hz, $\omega _{\bot }=10$kHz and $N=1000$ so that
$\gamma \sim 0.001$. The parameters of the system in the TG regime
are $\omega _x=10$Hz, $\omega _{\bot }=2000$kHz and $N=50$ with
$\gamma \sim 15$. The density profiles are shown in the Fig. 2 in
units of $N/a$ for m$=0$ and m$=0.2$, respectively. In the TG
regime they exhibit similar fermionization behavior just as in the
case of its ferromagnetic counterpart $^{87}$Rb. The only
difference is that here the population in 0 component remains
completely suppressed at zero magnetic field according to the mean
field theory. A straight explanation is that atoms prefer to be
aligned anti-parallel due to the anti-ferromagnetic spin
interaction. On the other hand, atoms in 0 component would collide
into pairs with one atom in + component and the other in -
component in order to lower the spin interaction energy. This
result is in agreement with the calculation for a spinor
condensate confined in a spherically symmetric 3D harmonic trap,
in which case the condensation occurs for + and - components
respectively \cite {WX Zhang2}.

\begin{figure}[tbp]
\includegraphics[width=3.22in]{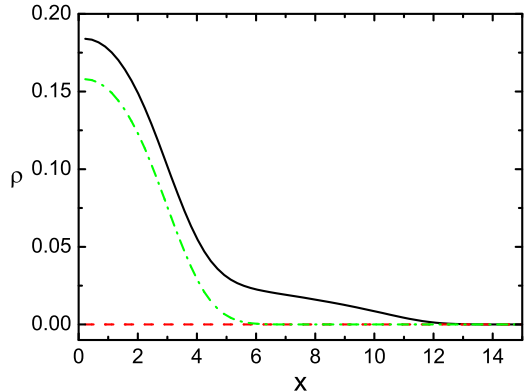}
\caption{(Color online) The density profiles of the spin-1
$^{23}$Na condensates in the ground state for the + (solid line),
0 (dashed line) and - (dash-dot lines) components respectively.
The interaction between atoms is enhanced such that spin
interaction parameter $g_2$ is at the same magnitude as that of
the density interaction $\tilde{g}_0$. In this figure the lengthes
are in units of 7.4 $\mu $m.} \label{fig3}
\end{figure}

For the condensates in the TG regime, the pair interaction
coefficient $ \tilde{g}_0=N^2\pi ^2/2$ is clearly a constant
irrespective of the $s$ -wave scattering length. Should the
interaction between atoms be enhanced by the so-called Feshbach
resonance, the pair interaction coefficient $g_2$ could be
increased greatly to the magnitude order of the coefficient
$\tilde{g}_0.$ This enable us to investigate explicitly the effect
of the spin-spin interaction. As an example, we consider the
system of $N=15$ $^{23}$Na atoms and m$=0.2$ in the harmonic trap
with $\omega _x=10$Hz and $\omega _{\bot }=2000$kHz. This
correspond to a spin interaction parameter $g_2=500$ while $
\tilde{g}_0=1110.33$ if the $s$-wave scattering length $a_2$ is
enhanced to $104.8a_B$. The corresponding density profiles are
given in Fig. 3. It is shown that the density profiles are no
longer Fermi-like and most atoms are compressed to the narrower
range around the center of the harmonic trap. A naive explanation
may be that pairs of atoms form singlets due to the strong
spin-spin interactions and the effective interaction between the
singlets is relatively very weak, therefore the picture of Tonks
gas breaks down. Finally, we discuss the conditions under which
the above result makes sense. For the $^{23}$Na atoms discussed
here, the spin exchange energy is always weak since $c_2$ is two
order of magnitude smaller than $c_0$. Even for the case in which
$a_2$ is enhanced to $104.8a_B$, we have $c_0/c_2=4.14$ and the
theory presented in this paper still holds. However, for a
condensate which enters the regime with $g_2$ larger than
$\tilde{g}_0$, our result is obviously not applicable.

\section{Summary}

The density profiles of 1D spin-$1$ Bose gases in the ground state
are evaluated in both the weakly interacting regime and the
strongly interacting TG regime. The population of atoms in
different components depends on the overall magnetization and the
(anti-)ferromagnetism of the Bose gases. When the system is in the
Tonks regime, the density profiles show obvious Fermi-like
distribution. However, for strong enough spin-spin interaction, we
observe apparent deviation of the density distribution from the
Fermi-like distribution in the TG regime.

\begin{acknowledgments}
The authors acknowledge the NSF of China (Grant No. 90203007 and
Grant No. 10574150) for financial support. SC is also supported by
the "Hundred Talent" program of Chinese Academy of Sciences. We
thank L. You, Y. Wang, W.-X. Zhang and W.-D. Li for useful
discussions.
\end{acknowledgments}

\end{document}